\documentclass{article}
\usepackage[T1]{fontenc} 
\usepackage[utf8]{inputenc} 
\usepackage{ismir,amsmath,cite,url}
\usepackage{graphicx}
\usepackage{color}
\pdfoutput=1

 \usepackage[bookmarks=false,hidelinks]{hyperref}

\usepackage{cleveref}
\usepackage{algorithm}
\usepackage{algpseudocode}
\usepackage{balance}

\newcommand{\setmeter}[2]{\ensuremath{%
\vcenter{\offinterlineskip
\halign{\hfil##\hfil\cr
$\scriptstyle#1$\cr
\noalign{\vskip1pt}
$\scriptstyle#2$\cr}
}}%
}
\newcommand{\muskern}{\kern-.15ex } 
\makeatletter
\newcommand\dynmark[1]{{\normalfont\bfseries\itshape
  \@tfor\next:=#1\do{\put@muskern\next}\/}}
\newcommand{\put@muskern}{\let\put@muskern\muskern}
\makeatother

\title{A Convolutional Approach to Melody Line Identification in Symbolic Scores}

\multauthor
{Federico Simonetta$^1$ \hspace{1cm} Carlos Cancino-Chac\'{o}n$^2$ \hspace{1cm} Stavros Ntalampiras$^1$} { \bfseries{Gerhard Widmer$^{3,2}$}\\
  $^1$ Music Informatics Laboratory, Dept. of Computer Science, University of Milano, Italy\\
$^2$ Austrian Research Institute for Artificial Intelligence, Vienna, Austria\\
$^3$  Dept. of Computational Perception, Johannes Kepler University Linz, Austria\\
{\tt\small federico.simonetta@unimi.it, carlos.cancino@ofai.at}
}
\def\authorname{Federico Simonetta, Carlos Cancino-Chac\'{o}n, Stavros Ntalampiras, Gerhard Widmer}

\begin{document}

\maketitle
\begin{abstract}
In many musical traditions, the melody line is of primary
significance in a piece. Human listeners can readily distinguish
melodies from accompaniment; however, making this distinction
given only the written score – i.e. without listening
to the music performed – can be a difficult task.
Solving this task
is of great importance for both Music Information Retrieval and
musicological applications.
In this paper, we propose an automated approach to
identifying the most salient melody line in a symbolic score.
The backbone of the method consists of a convolutional
neural network (CNN) estimating the probability that each
note in the score (more precisely: each pixel in a piano roll
encoding of the score) belongs to the  melody line.
We train and evaluate the method on various
datasets, using manual annotations where available and
solo instrument parts where not. We also propose a method
to inspect the CNN and to analyze the influence exerted by
notes on the prediction of other notes; this method
can be applied whenever the output of a neural network
has the same size as the input.

\end{abstract}

\section{Introduction}\label{sec:introduction}

Many musical traditions make use of melody-accompaniment structures. 
Generally, the melody line carries the most significant meaning, while the accompaniment provides harmonic and rhythmic support. 

In Western art music -- which, unlike music in some other traditions, is typically notated -- special attention is paid to the construction of melodies during composition.
Ideally, melodies in Western art music styles should involve an intervallic structure that is dependent on the specific tonal hierarchy defined by the piece~\cite{SalzerSchachter:1989,PistonCounterpoint:1947}.
Musicians typically accentuate melody lines during performance as a way of clarifying the piece structure for listeners: for example, melody lines may be played louder and with more flexible timing than accompaniment~\cite{doi:10.1121/1.1376133,Goebl:2016kl}.



Most listeners readily distinguish melody lines from accompaniment.
In contrast, identifying the melody line through visual inspection of a musical score -- without hearing the piece -- can be a difficult task, even for trained musicians \cite{Brodskyetal:2003}. 
In this paper, we propose a convolutional approach for identifying the melody line of a piece using a piano roll representation of the score. 
A solution for this task has potential implications for music information retrieval and musicology \cite{8665366}.
An effective algorithm could be applied to music retrieval tasks such as query-by-humming, searching a database of MIDI files for melodies, developing performance models that account for melody in predicting musical expression, etc.
Our focus is on music of the common practice period that uses melody-dominated homophonic textures (i.e., a single melody line plus accompaniment lines), rather than equal-voice polyphony (i.e., multiple independent melody lines) or monophony (i.e., unison melody shared by all voices). However, we provide extensive tests of the proposed method in styles other than common practice era, such as pop, baroque and contemporary art music.

The rest of this paper is structured as follows:
In Section \ref{sec:related_work}, we discuss related work on voice separation and streaming.
Section \ref{sec:baseline} briefly describes the baseline methods that we used for comparison against our model.
Section \ref{sec:method} presents a description of the proposed method.
Section \ref{sec:datasets} describes the three datasets used in this work.
Section \ref{sec:experiments} describes the experimental evaluation of the proposed method.
Section \ref{sec:results_and_discussion} discusses the results of the experimental evaluation.
Finally, Section \ref{sec:conclusions} concludes this paper and proposes some future research
directions. 
A companion website was also created to show additional material for the sake of reproducibility.\footnote{\label{websiteurl}\url{https://limunimi.github.io/Symbolic-Melody-Identification/}}

\begin{figure}[!t]
    \centering
    \includegraphics[width=0.45\textwidth]{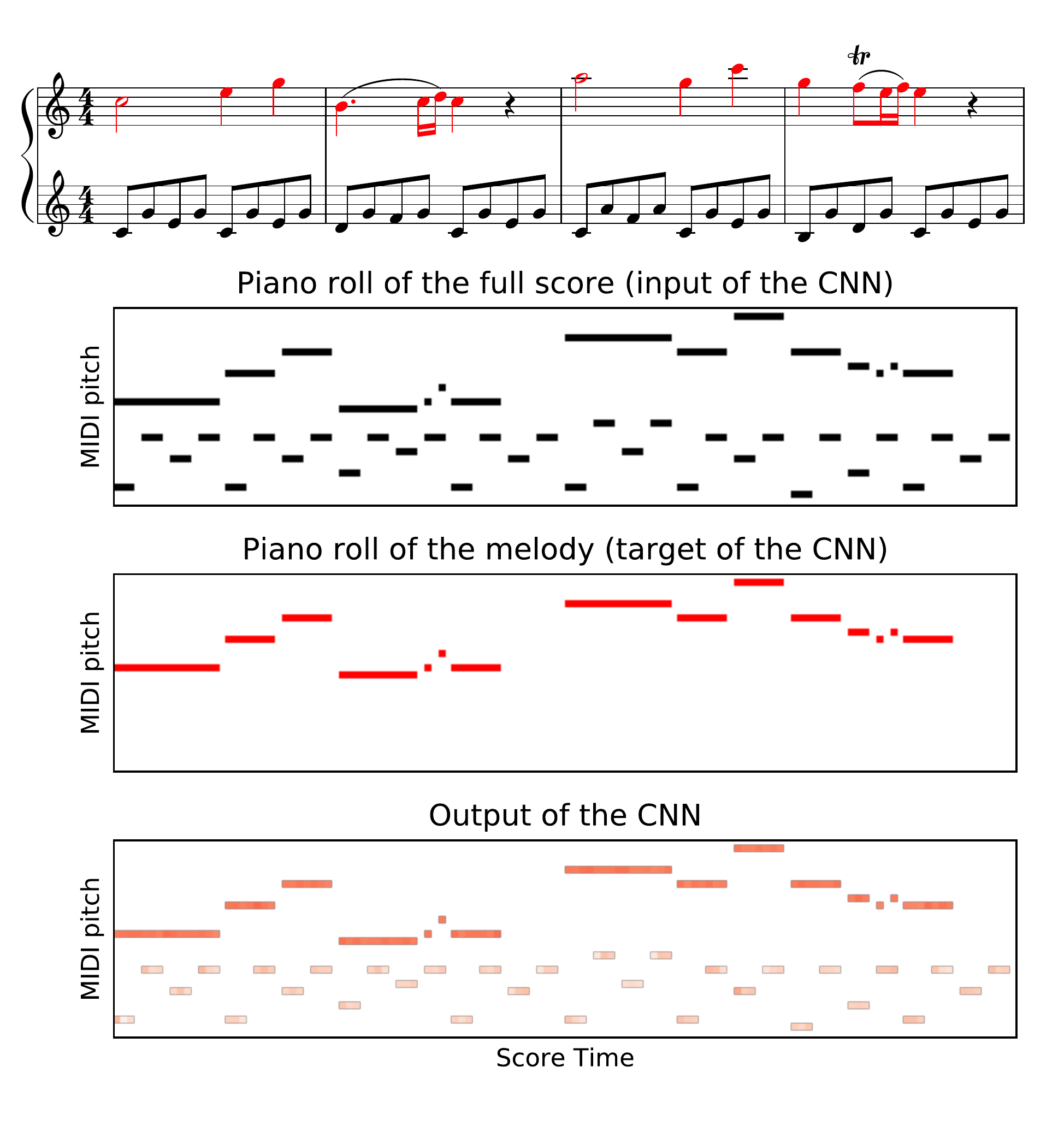}
    \caption{
    Top: Excerpt of Mozart's Sonata K. 545 (melody highlighted in red).
    Middle: Piano roll representation of the score (melody is highlighted in red).
    Bottom: Prediction of the CNN for this excerpt. In this piano roll, the intensity of the color of each pixel represents its probability of belonging to the melody.
    }
    \label{fig:piano_rolls}
\end{figure}

\section{Related Work}\label{sec:related_work}
\subsection{Voices and Streams}
\begin{figure*}[t]
    \centering
    \includegraphics[width=\textwidth]{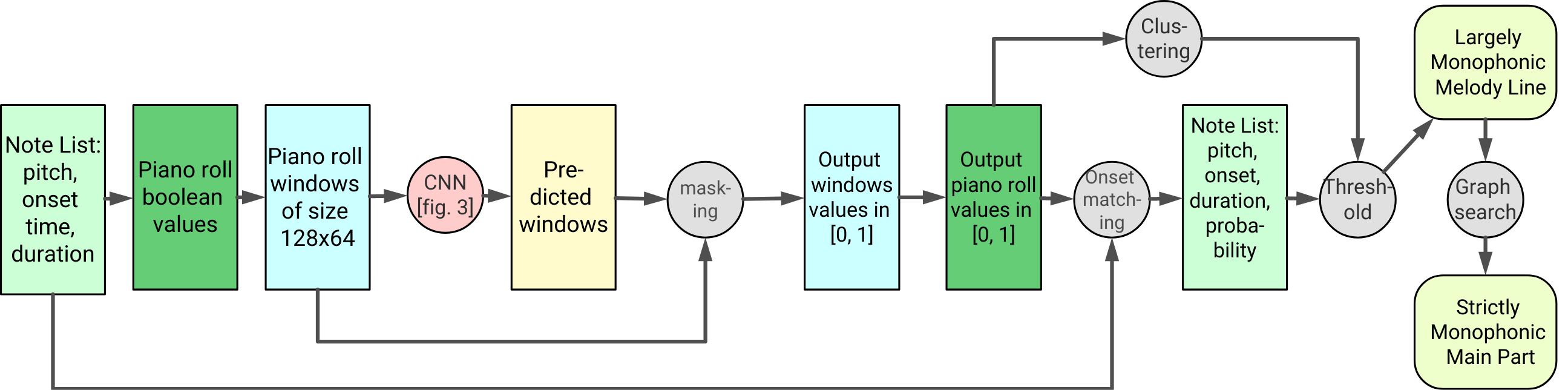}
    \caption{The pipeline of the proposed method (see Section \ref{sec:method}). 
    }
    \label{fig:schema}
\end{figure*}

 Music perception research has investigated listeners' abilities to distinguish between voices in homo- and polyphonic music, and has shown that the theoretical rules of voice leading are motivated by listeners' abilities to follow voices~\cite{Huron:2001ok}. 
Cambouropoulos~\cite{Cambouropoulos:2008pw} proposed three ways of defining musical ``voices'':
(1) for multi-instrument music, each instrument can be said to constitute a separate voice; this would allow for the possibility of non-monophonic voices in instruments that produce chords;
(2) voices can be assigned to melodic streams as they are perceived and segmented by listeners, following cognitive grouping principles; and
(3) in monophonic music, the harmonic content of the piece may imply a horizontal organization of polyphonic voices that unfold over time (i.e., implied polyphony), e.g., multiple temporally-overlapping voices could be assigned to passages of Bach's Cello Suites.
In this work, we use the second definition, and we define the melody line as the most salient voice.

In the music information retrieval literature, three corresponding tasks have been addressed: 
1) voice separation from symbolic scores~\cite{Gray2016, McLeod2016, Guiomard-Kagan2016, Chew2005}; 
2) main track identification (from MIDI files with multiple tracks)~\cite{HUANG2010,Martin2009, Li2009, Friberg2009}; and 
3) main melody identification from audio \cite{Bittneretal:2015pl,Bosch:2016ok,SalamonEtAl2018}. 
The latter is a different problem than that addressed here: it deals with the complex task of identifying notes from an audio file, but can use performance cues (e.g., contrasts in timbre and dynamics, which are not present in MIDI data) to facilitate melody identification.

Most relevant to the current study is the task of voice separation from symbolic scores. 
Some of the proposed methods are computational implementations that attempt to capture perceptual rules of segmentation \cite{Cambouropoulos:2008pw, Gray2016, Guiomardetal:2015, Guiomard-Kagan2016, Chew2005} -- in particular those rules codified by Huron~\cite{Huron:2001ok}.
For a more in-depth discussion on voice separation algorithms from symbolic scores, we refer the reader to \cite{Guiomardetal:2015,Isikhan2008,Wieringetal:2009,deValk:2018}.

\section{Baseline Methods}\label{sec:baseline}
\subsection{Skyline Algorithm}
The skyline algorithm is a heuristic that takes the highest note at each point in time \cite{Chai2001,Uitdenbogerd1998}. 
In Western art music, pop and many folk traditions from around the world, melodies are often carried by the highest voice.
After the submission of this paper, we discovered that a new method was being submitted for this same task \cite{Jiang2019}, confirming the relevance of this topic.

\subsection{VoSA}

Proposed by Chew and Wu~\cite{Chew2005}, VoSA is a successful voice separation method.
In this approach, a piece is split into segments based on voice entry and exit points, so that the number of sounding notes is constant within each segment.
The segment with the highest number of sounding notes defines the number of voices in the piece.
Notes are then connected into voices using connection weights, equal to the absolute size of the interval between one note and the next.
Like most voice separation methods, VoSA was designed to work with polyphonic rather than homophonic music.
In spite of its apparent simplicity, VoSA has been favorably compared against more sophisticated computational models of voice separation~\cite{Guiomardetal:2015, McLeod2016, Guiomard-Kagan2016}.


\section{Method}\label{sec:method}
\begin{figure}
    \centering
    \includegraphics[width=0.45\textwidth]{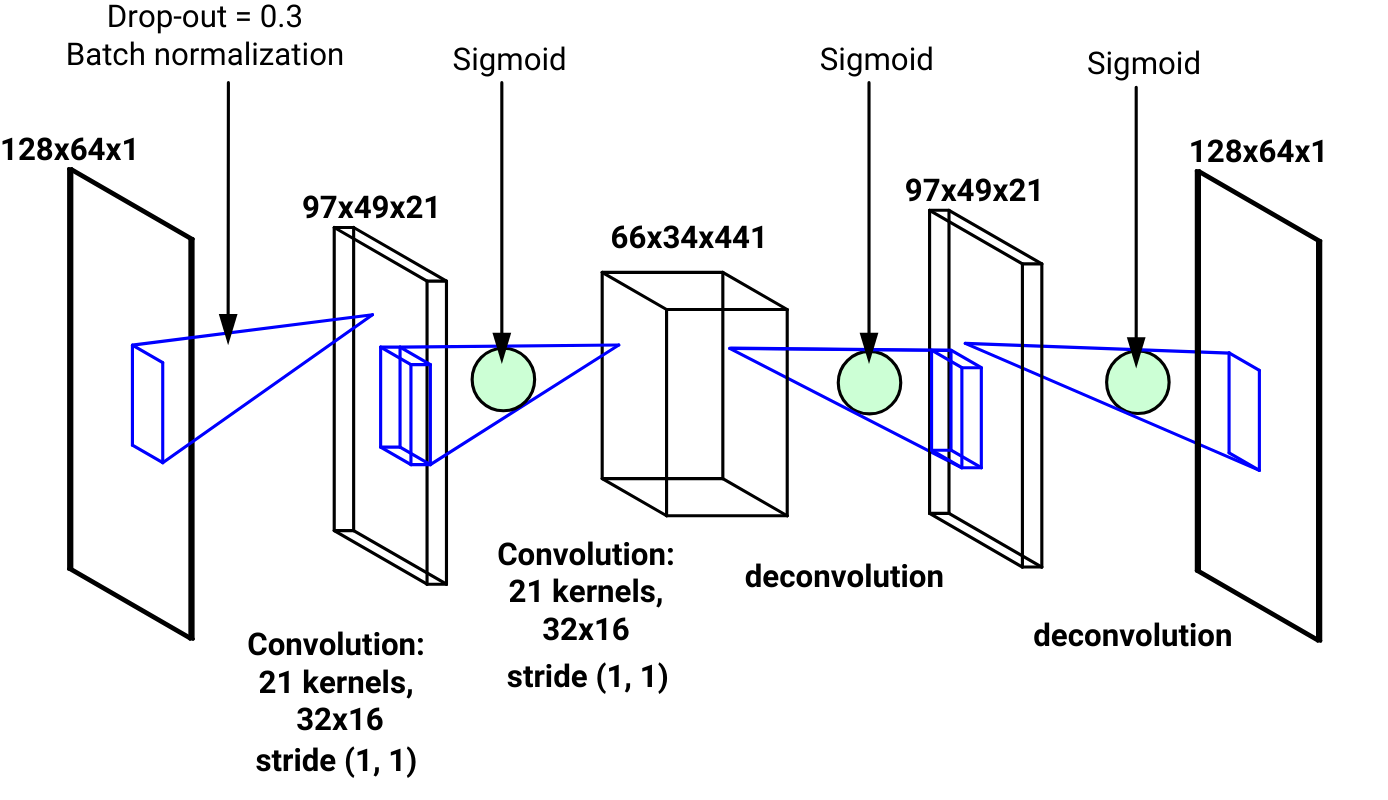}
    \caption{The architecture of the fully convolutional neural network used in the proposed method.
    The architecture of the network was determined using hyper-parameter optimization (see Section \ref{sec:architecture} for explanation). 
    }
    \label{fig:cnn}
\end{figure}

\subsection{Music Score Modeling Using CNNs}
\label{sec:CNN}
A schematic representation of our method is given in Figure \ref{fig:schema}. 
The backbone of the method consists of a fully convolutional neural network (shown in Figure \ref{fig:cnn}), which takes as input segments of a music score, represented as a piano roll, and estimates the probability that each note in the score (more precisely: each pixel in the piano roll encoding) belongs to the melody line.

A piano roll can be described as a 2D representation of a musical score; the x-axis indicates score time and the y-axis indicates pitch.
The piano rolls used in this study are constructed with a temporal resolution of 8 pixels/beat (i.e., a pixel represents a 32nd note in \setmeter{4}{4}). 
The piano roll of each piece is divided into overlapping fixed-length windows of 64 pixels (i.e., 8 beats). The length of the window was determined using hyper-parameter optimization, (see Section \ref{sec:architecture}).
The overlap between windows is 50\% (i.e., 2 beats), and windows shorter than this size are padded with zeros.
An output piano roll for each full piece is constructed by averaging probabilities for the pixels located in overlapping windows.
Afterwards, we apply a mask on the output piano roll by multiplying it by the (binary) input piano roll, so that areas with no notes take values of zero, and non-zero probabilities only remain where there are notes. 
The probability of each note belonging to the melody is then calculated as the median across the output values of its pixels.
In the following discussion, we will use \emph{note probability} as a shorthand to refer to the probability of a note to belong to the melody.


In Figure \ref{fig:piano_rolls} we show an excerpt of Mozart's Piano Sonata K.~545 and three vertically aligned piano rolls corresponding to the excerpt. 
The second row of this figure is the input piano roll, while the third row gives the ground truth melody line that we aim to identify in the input.
The bottom gives the piano roll that we obtain as output.
The output is color-coded with the notes that were identified as melody highlighted in red.


A threshold is needed to determine which note probabilities should indicate melody notes. 
Distributions of probabilities differ between pieces, so a hard threshold (e.g. $0.5$) would be inappropriate.
Instead, we find a threshold for each piece using a statistical analysis of the values of the note probabilities.
In the implementation of the proposed method we use hierarchical \textit{single-linkage} clustering \cite{Muellner2013}: two clusters across the values of note probabilities are identified, and a piece-wise threshold is selected as the largest value of the lowest cluster.
We then compare each note probability to this threshold and either retain the note as melody or filter it out as accompaniment.
This produces largely (but not entirely) monophonic melody output -- in some cases, multiple simultaneous notes pass the threshold.
A graph-based method, explained next, was thus implemented to select a strictly monophonic melody line from this output.

\subsection{Graph Search}\label{sec:graph}

\begin{figure}
    \centering
    \includegraphics[width=0.5\textwidth]{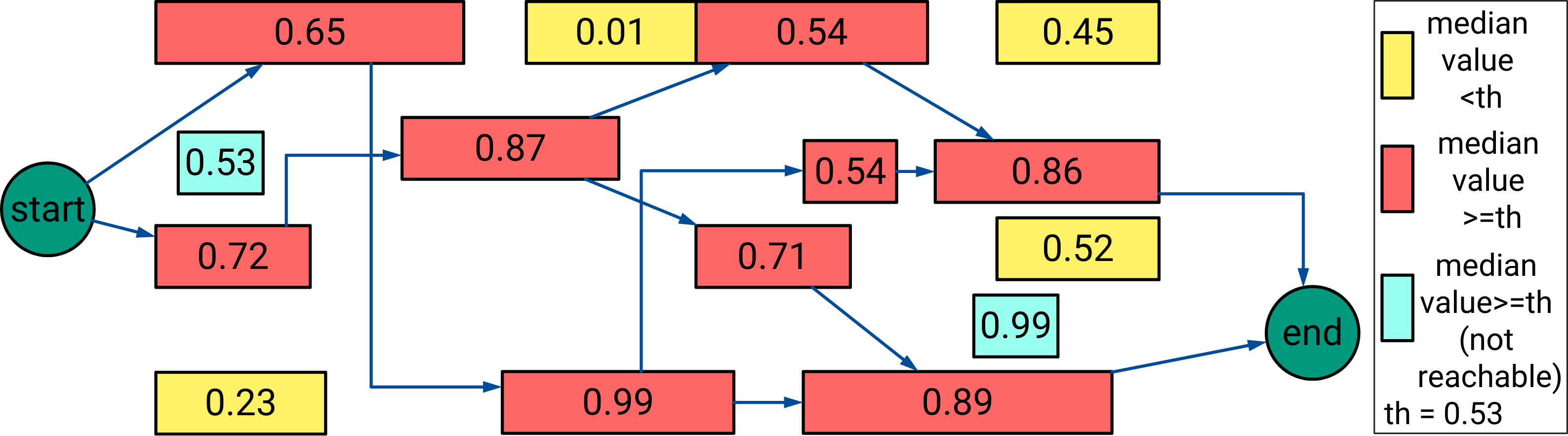}
    \caption{Example of graph built with \Cref{alg:graph}. Red notes are notes over
        threshold, yellow notes are under threshold, while blue notes are over threshold but
        are not reached by any path. The green circles are the starting and ending nodes.
        Numbers indicate note probabilities, which are computed as the median of their pixels.
    }
    \label{fig:graph}
\end{figure}

Having identified notes that pass the threshold as defined above, we have to select a sequence of these notes that maximizes the probability of the sequence being monophonic. 
This is achieved using a graph-based approach. 
Algorithm \ref{alg:graph} is used to build a \emph{directed acyclic graph} (or digraph, see Figure \ref{fig:graph}).
Such a graph consists of a set of nodes and a set of directed edges.
Each of these edges specifies a connection from a node to another.
In the graph defined by Algorithm \ref{alg:graph}, each note that passes the threshold is represented by a node, and the pitch, onset and duration information of this note are used to determine to which nodes is the note connected (in order to guarantee a strictly monophonic sequence).
Note probabilities are used to determine the strength of the connection between nodes (similar to a ``distance''; notes with high probabilities are considered ``closer'').
Additionally, we set a start and end node at the beginning and end of the piece, respectively. 
We can then use a single-source shortest path algorithm to find the main melody line as the shortest path from the start to the end nodes.
In our current implementation, we use the negative note probabilities as connection weights and the Bellman-Ford algorithm\footnote{\url{https://docs.scipy.org/doc/scipy-1.2.1/reference/generated/scipy.sparse.csgraph.bellman_ford.html}} to find the shortest path through the graph.\footnote{Depending on the choice of the connection weights, other shortest path algorithms (e.g., topological sorting, Dijkstra's, etc.) are possible.}

\begin{algorithm}[t]
    \caption{Melo-digraph building}\label{alg:graph}
    \begin{algorithmic}
        \State $L \gets $ list of notes
        \State $\alpha \gets $ starting node (end time $=0$)
        \State $\omega \gets $ ending node (onset $=\infty$, probability $=-0.5$)
        \State Push $\alpha$ to the beginning of $L$
        \State Push $\omega$ to the end of $L$
        \For{$note$ in $L$}
        \State $L' \gets $ notes with onset $\ge$ end time of $note$
        \State $L' \gets $ notes with onset $=$ minimum onset in $L'$
        \For{$note'$ in $L'$}
        \If{probability of $note'$ $\ge $ threshold}
        \State $p = $ probability of $note'$
        \State add an edge $(note, note')$ with weight $-p$
        \EndIf
        \EndFor
        \EndFor
    \end{algorithmic}
\end{algorithm}

\subsection{Training}
The CNN is trained in a supervised fashion to filter out accompaniment parts. 
Inputs are provided in the form of piano roll segments and the targets are the corresponding piano rolls with only the melody notes. 
We also augmented the training dataset by $50\%$ by creating copies of the original examples in the dataset with the melody transposed down for 2 octaves or up for 1 octave.
Though the standard loss function for binary classification problems like this one is the binary cross entropy, during development of the model, we achieved more accurate models by minimizing mean squared error for the match between output and target piano rolls.
The networks were trained using AdaDelta\cite{zeiler2012adadelta} with initial learning rate set to $1$.
In order to avoid overfitting, we use dropout with probability $p_{dropout}=0.3$ and $L_1$-norm weight regularization.
Additionally we use batch-normalization~\cite{ioffe2015batch}.
The training is stopped after 20 epochs without improvement in validation loss~\cite{nips1989_275} .

\begin{figure*}
    \centering
    \includegraphics[width=0.95\textwidth]{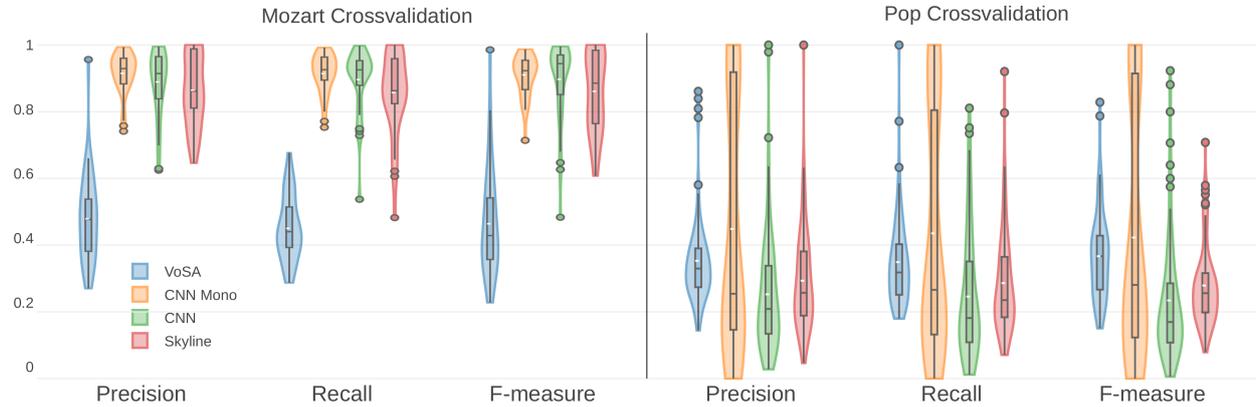}
    \caption{Cross-validation on Mozart and Pop datasets. With the Wilcoxon test applied
        to F-measure, we found a significant difference between \textit{CNN Mono} and
        \textit{VoSA} and between \textit{CNN Mono} and \textit{CNN}, but no
        significant difference was found between \textit{CNN Mono} and \textit{Skyline} in the Pop
        dataset (only in the Mozart dataset). The mean is marked with a white dash.
    }
    \label{fig:crossvalidation}
\end{figure*}

\begin{figure}
    \centering
    \includegraphics[width=0.5\textwidth]{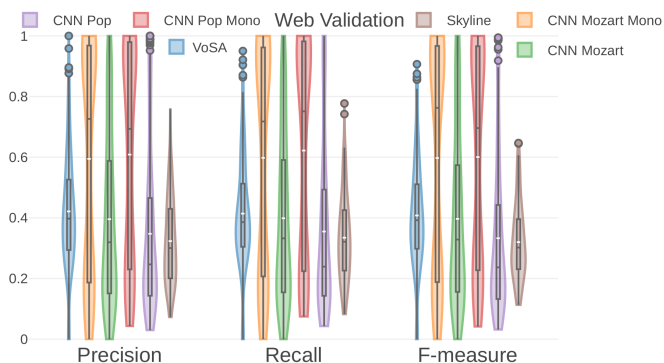}
    \caption{Validation on the Web music dataset. With the Wilcoxon test, we found
        a significant difference between \textit{Mono} models and \textit{Skyline}/\textit{VoSA},
        but there was not always a significant difference when comparing non-\textit{Mono} models
        and \textit{Skyline}/\textit{VoSA}.  The mean is marked with a white dash.
    }
    \label{fig:validation}
\end{figure}

\section{Datasets}\label{sec:datasets}
We used three different datasets to evaluate the performances of our method.
The first dataset (``Mozart'') consists of $38$ movements from (13) Mozart Piano Sonatas, for which the main melody line was annotated manually by a professional pianist.
The second dataset (``Pop'') consists of $83$ popular songs (including pop and jazz).
We used the vocal part of these songs as the melody line, and treated them as though they were compressed onto a single track (identifying the main track in multi-track music is a separate question, see~\cite{HUANG2010,Martin2009, Li2009, Friberg2009}).

These datasets were used for training and testing.
A third dataset (``Web''), used only for testing, comprises MIDI files crawled from the web.
This dataset includes $169$ Western art music compositions from the late 16th to the early 20th centuries.
All of these pieces included a solo instrument (typically voice, flute, violin or clarinet) and accompaniment (typically strings or piano).

The first and third of these datasets are publicly available for research purposes in the companion site -- see footnote 1.
We do not have distribution rights for the second dataset, which was professionally curated and annotated, but we provide the full list of pieces.





\section{Experiments}\label{sec:experiments}
\subsection{Evaluation Metrics and Baseline Methods}
In all experiments, we evaluated the quality of the predictions using the F-measure.
We experimented on the largely monophonic (which we denote \emph{cnn} in the following discussion) and strictly monophonic (denoted as \emph{cnn mono}) variants of the proposed model described in Sections \ref{sec:CNN} and \ref{sec:graph}, respectively.
As a baseline comparison, we used the skyline algorithm and VoSA (both described in Section \ref{sec:baseline}).
Since VoSA does not directly output the melody line, we first separate the piece into individual voices (as identified by VoSA), then select the voice with the highest F-measure as the melody.
These modifications allowed us to consider the best case scenario of VoSA.

\subsection{Network Architecture}\label{sec:architecture}
To determine the architecture of the network, we used hyper-parameter optimization.\footnote{Using the ``hyperopt'' library in Python~(\url{http://hyperopt.github.io/hyperopt/}).}
The number of convolutional layers, kernel size and number, and window lengths were optimized. 
This hyper-parameter optimization was done on $100$ pieces randomly selected from across the three datasets plus $65$ MIDI files collected online using the same criteria as the Web dataset.
To compare models, we constructed training, validation, and test sets from the 100 pieces.
A model configuration was selected that performed most successfully on the test set.
The selected network architecture is shown in Figure \ref{fig:cnn}: 
$2$ convolutional layers, each with $21$ kernels of size $32\times16$ (i.e., over two and a half octaves in the pitch dimension and 2 beats in the time dimension).

\subsection{Evaluation of the Proposed Method}\label{sec:predictive_accuracy}

To evaluate the quality of the predictions of the proposed method, we conducted two experiments.
In the first experiment, we were interested in evaluating the predictive accuracy of the models trained on different datasets.
In the second experiment we tested how well models generalize to different music styles.
For the first experiment, we performed a 10-fold cross-validation on each of the Mozart and Pop datasets. 
In each of these cross-validations, the dataset was split into 10 folds.
The model was trained on 9 of these folds and tested on the remaining one. 
We did this for all possible combinations so that each piece in each dataset appeared in the test set once.
For the second experiment, we tested models trained on Mozart and models trained on Pop on the Web dataset.

\begin{figure*}
    \centering
    \includegraphics[width=0.95\textwidth]{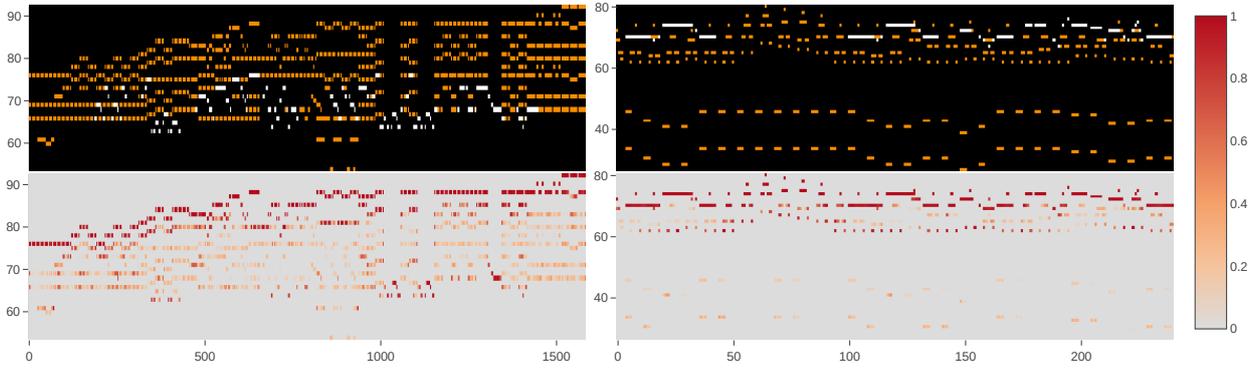}
    \caption{Liszt's \textit{Ihr Glocken von Marling} (left) and an excerpt from
        Schubert's \textit{Ave Maria} (right). Input piano roll (above), prediction of the
        CNN (middle). In Liszt, the
        model fails to identify the main part because the texture is rather different
        from the most common case and the melody is in the middle voices. In Schubert,
        instead, the texture changes but the model is not able to identify when the main
        part starts and stops because the accompaniment plays similar notes.
    }
    \label{fig:schubert-liszt}
\end{figure*}

\section{Results and Discussion}\label{sec:results_and_discussion}
\subsection{Model Performance}
The violin plots summarizing the results of these experiments are shown in Figures \ref{fig:crossvalidation} and \ref{fig:validation}, while detailed plots are available in the companion website (see footnote 1). 

Our first experiment tested how well models predicted melody lines given training and testing on the same genre of music. 
Wilcoxon signed-rank tests were run on F-measures to assess potential differences between models.
Test results are described in the caption of Figure \ref{fig:crossvalidation}.
Overall, our proposed method that identified strictly monophonic melody lines (\emph{cnn mono}) performed better than the other models, but this difference was only significant for the Mozart dataset.
The Mozart pieces are highly structured and their melody lines tend to occur in the upper-most voice. 
The Pop dataset, in contrast, contains pieces with variable structure, with longer breaks in the melody (e.g., there is sometimes an interlude in the accompaniment part).
Furthermore, the accompaniment part often overlaps in register with the melody line.
It seems that without additional timbral information, our model could not sufficiently distinguish between melody and accompaniment lines when they shared a similar texture.

Our second experiment tested how well trained models generalize to new types of data (i.e., Web dataset).
We hypothesized that models trained on the Mozart dataset would outperform models trained on the Pop dataset, as the Mozart and Web datasets are more similar in style (though the Web dataset is more heterogeneous). 
However, no significant difference between models was found -- both models performed well on the Web dataset.

Regarding the less-successful performance of the two baseline methods, the skyline method fails when the melody is not the highest voice; furthermore, this method cannot identify when pauses occur in the solo part. 
The VoSA method, which was developed for use with polyphonic music, tends to create too many voices and shows a bias towards connecting notes separated by small intervals -- this is not surprising, as polyphonic music tends to assign voices to small pitch ranges.
As a result, accompaniment notes are often wrongly included in the melody line that VoSA identifies.

\subsection{Saliency Maps}
To investigate what the CNNs are learning, we propose a method (similar to a sensitivity analysis) that evaluates the contribution of individual locations of the piano roll to predictions at other locations using saliency maps.\footnote{Kernels, saliency maps and additional material are available on the companion website -- see footnote 1.}
The method involves testing how the probability that a given note belongs to the melody changes (i.e., increases or decreases) when certain other notes are removed (i.e., by converting the pixels belonging to those notes to 0).

For example, take a rectangular input window $I$ and its prediction $P$. A new input window $I'$ with prediction $P'$ is created by converting the pixels inside a given rectangle $R$ to 0. 
The difference between the original and new predictions is denoted as $d(P, P')$ and can be interpreted as the contribution given by the notes inside $R$ to the original prediction.
By testing different input windows across the piano roll, we can see how different elements of the music contribute to the predictions that are obtained for individual notes.

If we are interested in a particular note $n$, we can compute $d(P, P')$ specifically for the pixels belonging to $n$.
For our analysis, for certain notes of interest, we define 5 randomly-positioned rectangles $R$ and calculate $d(P, P')$. This difference is summed to the pixels of the notes inside each rectangle $R$.
This procedure is repeated $N$ times (where $N$ is a trade-off between computational complexity and resolution of the saliency map; in our case $N=30000$), and we select only the iterations in which the pixels of note $n$ are not converted to 0. 
Each pixel is then normalized by the number of times it was converted to 0. 
As difference we use

\begin{equation}
d(P, P') =  \frac{\sum_{i=n_{start}}^{i=n_{end}}{P[i] - P'[i]}}{Area(n_{start}, n_{end})} \end{equation}
where $n_{start}$ and $n_{end}$ identify the region occupied by the note $n$.
In general, $n_{start}$ and $n_{end}$ indicate two opposite corners of any rectangle.

With this difference function, given a rectangle $R$, if $d(P, P')>0$, then $P>P'$ in average across $n$ and, thus, removing the notes inside $R$ decreases the prediction values of $n$; conversely, if $d(P, P')<0$, then $P<P'$ and removing the notes inside $R$ increases the prediction.

\begin{figure}[!hb]
    \centering
    \includegraphics[width=0.5\textwidth]{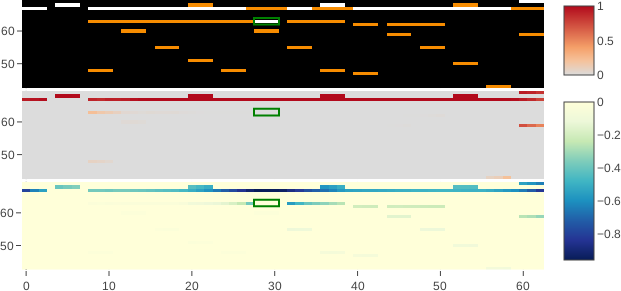}
    \caption{Input piano roll with ground truth in white (top), prediction of the CNN (middle) and proposed saliency computed with respect to the green rectangle (bottom).}
    \label{fig:saliency}
\end{figure}

For example, in the bottom piano roll in Figure \ref{fig:saliency}, the blue high-pitched notes occurring around beats 20 and 35 have non-positive saliency values.
Because they are higher pitched, these notes contribute negatively to the melody note highlighted with a green box, making it unlikely for this note to be identified as melody.
In the companion website, we show the saliency of other regions highlighting that the prediction of some notes is influenced positively by some regions and negatively by others and that the CNN exploits the regular patterns in the accompaniment to identify the melody notes.

Overall, our model incorporates features of both the skyline algorithm and VoSA.
Like the skyline algorithm it focuses on the highest notes of the piece; on the other hand, by allowing for different probabilities like VoSA, it is more successful at drawing coherent melody lines.
Unlike VoSA, however, our model does not incorporate explicit perceptual constraints.

\section{Conclusions}\label{sec:conclusions}
We implemented and analyzed a novel method to identify the melody line in a symbolic music score. 
Some of the functions of our model were found to be similar to functions of the skyline algorithm and VoSA (in particular, focusing on the upper-most pitch, and defining a melody line as finding the sequence of notes that minimizes the connection cost).
However, our method does not take into account the long-term sequential nature of music; it can compute windows in any order.
While such a property might have some practical benefits, it also makes the network unable to generalize to diverse textures, leading to poor results when musical texture is varied (e.g., Figure \ref{fig:schubert-liszt}).

The next step for this line of research would be to develop a model that can take into account a larger temporal context.
A promising approach would be to incorporate attention mechanisms into the network.



\section{Acknowledgements}
This research has received funding from the European Research Council (ERC) under the European Union's Horizon 2020 research and innovation programme under grant agreement No. 670035 (project ``Con Espressione").
We gratefully acknowledge the support of NVIDIA Corporation with the donation of the Titan V GPU used for this research.
We thank Elaine Chew for sharing the code for VoSA. We thank Laura Bishop for proofreading an earlier version of this manuscript.

\balance
\bibliography{bibliography}

\end{document}